\documentclass[a4paper,12pt]{article}
\usepackage{epsfig}
\usepackage{rotating}
\DeclareGraphicsExtensions{.eps,.eps.gz,.ps,.ps.gz}
\begin{document}
\newcommand{\NIMA}[3]{Nucl.\ Instrum.\ Methods {\bf A{#1}} ({#2}) {#3} }
\newcommand{\PLB}[3]{Phys.\ Lett.\ {\bf B{#1}} ({#2}) {#3}}
\newcommand{\PRL}[3]{Phys.\ Rev.\ Lett.\ {\bf {#1}} ({#2}) {#3}}
\newcommand{\PRD}[3]{Phys.\ Rev.\ {\bf D{#1}} ({#2}) {#3}}
\newcommand{\NP}[3]{Nucl.\ Phys.\ {\bf {#1}} ({#2}) {#3}}
\newcommand{\NPB}[3]{Nucl.\ Phys.\ {\bf B{#1}} ({#2}) {#3}}
\newcommand{\PR}[3]{Phys.\ Rep.\ {\bf {#1}} ({#2}) {#3}}
\newcommand{\ZPC}[3]{Z.\ Phys.\ {\bf C{#1}} ({#2}) {#3}}
\newcommand{\EPJC}[3]{Eur.\ Phys.\ J.\ {\bf C{#1}} ({#2}) {#3}}
\newcommand{\CPC}[3]{Comp.\ Phys.\ Comm.\ {\bf {#1}} ({#2}) {#3}}
\newcommand{\etal}{\mbox{\it{et~al.}}}
%
\newcommand{\BR}{\mbox{$\mathrm{BR}$}}
\newcommand{\X}{\mbox{$\chi^0$}}
\newcommand{\XD}{\mbox{$\chi^{0\prime}$}}
\newcommand{\mX}{\mbox{$m_{\chi^0}$}}
\newcommand{\mXD}{\mbox{$m_{\chi^{0\prime}}$}}
\newcommand{\Co}{\mbox{$\tilde{\chi}^0_1$}}
\newcommand{\Ct}{\mbox{$\tilde{\chi}^0_2$}}
\newcommand{\mCo}{\mbox{$m_{\tilde{\chi}^0_1}$}}
\newcommand{\mCt}{\mbox{$m_{\tilde{\chi}^0_2}$}}
\newcommand{\wwln}{\mbox{$\mathrm{W}^{+}\mathrm{W}^{-}
    \to\mathrm{q}\bar{{\mathrm{q}}^{\prime}}\lnu$}}
\newcommand{\ellell}{\mbox{$\ell^{+}\ell^{-}$}}
\newcommand{\sqrts}{\mbox{$\sqrt{s}$}}
\newcommand{\lnu}{\mbox{$\ell\nu$}}
\newcommand{\nn}{\mbox{$\nu\bar{\nu}$}}
\newcommand{\Zho}{\mbox{$\mathrm{Zh^0}$}}
\newcommand{\ZZ}{\mbox{$\mathrm{Z^0Z^0}$}}
\newcommand{\WW}{\mbox{$\mathrm{W^+W^-}$}}
\newcommand{\YTT}{\mbox{$y_{\mathrm{23}}$}}
\newcommand{\DTT}{\mbox{$d_{\mathrm{23}}$}}
\newcommand{\COST}{\mbox{$\cos\theta$}}
\newcommand{\ACOST}{\mbox{$|\cos\theta|$}}
\newcommand{\NGCH}{\mbox{$N^{\mathrm{good}}_{\mathrm{ch}}$}}
\newcommand{\NCH}{\mbox{$N_{\mathrm{ch}}$}}
\newcommand{\MVIS}{\mbox{$M_{\mathrm{vis}}$}}
\newcommand{\EVIS}{\mbox{$E_{\mathrm{vis}}$}} 
\newcommand{\EVISSQ}{\mbox{$E_{\mathrm{vis}}^{2}$}}
\newcommand{\EVISCT}{\mbox{$E_{\mathrm{vis}}^{\ACOST>0.9}$}}
\newcommand{\MMIS}{\mbox{$M_{\mathrm{miss}}$}}
\newcommand{\MMISSQ}{\mbox{$M_{\mathrm{miss}}^2$}}
\newcommand{\TMIS}{\mbox{$\theta_{\mathrm{miss}}$}}
\newcommand{\CMIS}{\mbox{$\cos\TMIS$}}
\newcommand{\ACMIS}{\mbox{$|\cos\TMIS|$}}
\newcommand{\TJET}{\mbox{$\theta_{\mathrm{jet}}$}}
\newcommand{\CJET}{\mbox{$\cos\TJET$}}
\newcommand{\ACJET}{\mbox{$|\cos\TJET|$}}
\newcommand{\MACJET}{\mbox{$\max(\ACJET)$}}
\newcommand{\acop}{\mbox{$\phi_{\mathrm{acop}}$}}
\newcommand{\ycut}{\mbox{$y_{\mathrm{cut}}$}}
\newcommand{\dm}{\mbox{$\Delta M$}}
\newcommand{\EBEAM}{\mbox{$E_{\mathrm{beam}}$}}
\newcommand{\Zo}{\mbox{$\mathrm{Z}^0$}}
\newcommand{\ee}{\mbox{$\protect \mathrm{e^{+}e^{-}}$}}
\newcommand{\rmee}{\mbox{$\mathrm{e^{+}e^{-}}$}}
\newcommand{\mm}{\mbox{$\protect \mu^{+}\mu^{-}$}}
\newcommand{\tautau}{\mbox{$\protect \tau^{+}\tau^{-}$}}
\newcommand{\ho}{\mbox{$\mathrm{h}^{0}$}}
\newcommand{\Ho}{\mbox{$\mathrm{H}^{0}$}}
\newcommand{\HoSM}{\mbox{$\mathrm{H}^{0}_{\mathrm{SM}}$}}
\newcommand{\qq}{\mbox{$\protect{\mathrm q}\protect\bar{\mathrm q}$}}
\newcommand{\mZ}{\mbox{$m_{\mathrm{Z^0}}$}}
\newcommand{\mZSQ}{\mbox{$m_{\mathrm{Z^0}}^2$}}
\newcommand{\Zs}{\mbox{${\mathrm{Z}}^{*}$}}
\newcommand{\gs}{\mbox{$\gamma^{*}$}}
\newcommand{\Zgs}{\mbox{${\mathrm{Z}/\gamma^{*}}$}}
\newcommand{\gamgam}{\mbox{$\gamma\gamma$}}
\newcommand{\clb}{\mbox{$1-\mathrm{CL_b}$}}
\newcommand{\pb}{\mbox{$\mathrm{pb}^{-1}$}}
\newcommand{\gev}{\mbox{$\mathrm{GeV}$}}
\newcommand{\mh}{\mbox{$m_{\mathrm{h^0}}$}}
\newcommand{\jj}{2-jet}
\newcommand{\jjj}{\mbox{$>$2}-jet}
\renewcommand{\thefootnote}{\alph{footnote}}
\newcommand{\pt}{\mbox{$p_{\mathrm{T}}$}}
\newcommand{\njets}{\mbox{$N_{\mathrm{jets}}$}}
\def\VERSION{1.5}
\def\luma{56.1}     
\def\lumb{178.2}    
\def\lumc{29.0}     
\def\lumd{71.7}     
\def\lume{74.9}     
\def\lumf{39.3}     
\def\lumg{6.3}      
\def\lumh{71.4}     
\def\lumi{124.6}    
\def\lumj{7.8}      
\def\lumtot{659.3}  
\def\ecma{182.7}   
\def\ecmb{188.6}   
\def\ecmc{191.6}   
\def\ecmd{195.5}   
\def\ecme{199.5}   
\def\ecmf{201.7}   
\def\ecmg{203.7}   
\def\ecmh{205.0}   
\def\ecmi{206.5}   
\def\ecmj{208.0}   
\def\ocra{4.1}
\def\ocrb{2.5}
\def\ocrc{2.5} 
\def\ocrd{2.2}
\def\ocre{2.1}
\def\ocrf{3.4}
\def\ocrg{2.2}
\def\ocrh{3.0}
\def\ocri{2.8}
\def\ocrj{2.5}
\def\mhsampA{105}
\def\limit{108.2}
\def\limie{108.6}
\def\limna{108.4} 
\def\limnb{107.0} 
\def\limnae{108.2}
\def\limnbe{107.3}
\begin{titlepage}
  \begin{center}
    \large{EUROPEAN ORGANIZATION FOR NUCLEAR RESEARCH}
  \end{center}
  \begin{flushright}
       CERN-PH-EP/2007-018 \\
       OPAL PR421 \\  
       12th June 2007  \\
  \end{flushright}
  \begin{boldmath}
    \begin{center}{\LARGE\bf Search for invisibly decaying Higgs bosons\\
	in ${\protect \boldmath \rmee\rightarrow\Zo\ho}$ production\\ 
	at $\sqrt{s}$ = 183 -- 209~GeV}
    \end{center}
  \end{boldmath}
\bigskip\bigskip
  \begin{center}{\LARGE The OPAL Collaboration}\end{center}
\bigskip\bigskip\bigskip
  \begin{abstract}
    A search is performed for Higgs bosons decaying into invisible
    final states, produced in association with a \Zo\ boson in \rmee\
    collisions at energies between 183 and 209 \gev. The search is
    based on data samples collected by the OPAL detector at LEP
    corresponding to an integrated luminosity of about 660 \pb. The
    analysis aims to select events containing the hadronic decay
    products of the \Zo\ boson and large missing momentum, as expected
    from Higgs boson decay into a pair of stable weakly interacting
    neutral particles, such as the lightest neutralino in the Minimal
    Supersymmetric Standard Model. The same analysis is applied to a 
    search for nearly invisible Higgs boson cascade decays into stable weakly
    interacting neutral particles. No excess over the expected
    background from Standard Model processes is observed.  Limits on
    the production of invisibly decaying Higgs bosons produced in
    association with a \Zo\ boson are derived.  Assuming a branching
    ratio $\BR(\ho\to\mathrm{invisible})=1$, a lower limit of
    \limit~\gev\ is placed on the Higgs boson mass at the 95\% confidence
    level. Limits on the production of nearly invisibly decaying Higgs
    bosons are also obtained.
  \end{abstract}
\bigskip\bigskip
\bigskip\bigskip\bigskip\bigskip
\begin{center}{\large
(Submitted to Physics Letters B)
}\end{center}
  \thispagestyle{empty}
\end{titlepage}
\begin{center}{\Large        The OPAL Collaboration
}\end{center}\bigskip
\begin{center}{
G.\thinspace Abbiendi$^{  2}$,
C.\thinspace Ainsley$^{  5}$,
P.F.\thinspace {\AA}kesson$^{  7}$,
G.\thinspace Alexander$^{ 21}$,
G.\thinspace Anagnostou$^{  1}$,
K.J.\thinspace Anderson$^{  8}$,
S.\thinspace Asai$^{ 22}$,
D.\thinspace Axen$^{ 26}$,
I.\thinspace Bailey$^{ 25}$,
E.\thinspace Barberio$^{  7,   p}$
% ...
}\end{center}\bigskip
\bigskip
$^{  1}$School of Physics and Astronomy, University of Birmingham,
Birmingham B15 2TT, UK
\newline
$^{  2}$Dipartimento di Fisica dell' Universit\`a di Bologna and INFN,
I-40126 Bologna, Italy
\newline
% ...
$^{  5}$Cavendish Laboratory, Cambridge CB3 0HE, UK
\newline
% ...
$^{  7}$CERN, European Organisation for Nuclear Research,
CH-1211 Geneva 23, Switzerland
\newline
$^{  8}$Enrico Fermi Institute and Department of Physics,
University of Chicago, Chicago IL 60637, USA
\newline
% ...
$^{ 21}$Department of Physics and Astronomy, Tel Aviv University,
Tel Aviv 69978, Israel
\newline
$^{ 22}$International Centre for Elementary Particle Physics and
Department of Physics, University of Tokyo, Tokyo 113-0033, and
Kobe University, Kobe 657-8501, Japan
\newline
% ...
$^{ 25}$University of Victoria, Department of Physics, P O Box 3055,
Victoria BC V8W 3P6, Canada
\newline
$^{ 26}$University of British Columbia, Department of Physics,
Vancouver BC V6T 1Z1, Canada
\newline
% ...
$^{  p}$ now at The University of Melbourne, Victoria, Australia
\newline
% ...
\bigskip
\section {Introduction}
The Higgs boson~\cite{Higgs} is required by the Standard Model
(SM)~\cite{sm} but has not yet been observed~\cite{Higgs-searches}.
At LEP II energies it should be produced mainly through the
``Higgs-strahlung'' process ($\rmee\to\Zs\to\Ho\Zo$) if its mass is
sufficiently low.  In the SM, the Higgs boson dominantly decays into a
pair of the heaviest kinematically accessible particles, which would
be a b-quark pair at LEP II.  In some models beyond the SM, however,
the Higgs boson can decay predominantly into a pair of invisible
particles if the process is kinematically allowed.

The Minimal Supersymmetric Standard Model (MSSM)~\cite{mssm} is one of
the models which allows for invisibly decaying Higgs
bosons~\cite{MSSM-inv}, through the $\ho\to\Co\Co$ process, where
\Co\ is the lightest neutralino, if the mass of \Co\ is lighter than
half of the Higgs mass and R-parity is conserved.  If \Co\ is purely
photino-like, the decay $\ho\to\Co\Co$ is suppressed.  In this case a
decay $\ho\to\Co\Ct$, where \Ct\ is the second lightest neutralino,
becomes dominant if it is allowed kinematically.  If the mass
difference (\dm) between \Ct\ and \Co\ is small, the visible products
of the decay $\Ct\to\Co\Zs/\gamma$ are soft and the event topology is
similar to that produced by an invisible Higgs decay $\ho\to\Co\Co$.
The $\ho\to\Co\Ct$ processes are therefore referred to as nearly
invisible Higgs decays.

In a non-linear supersymmetric model, the Higgs boson can decay into a
neutrino plus a Goldstino~\cite{goldstino}, and the invisible decay can
be dominant.
In other models beyond the SM with a spontaneously broken global
symmetry, the Higgs boson could decay into a pair of massless
Goldstone bosons, called Majorons~\cite{majoron}, which couple
strongly to the Higgs boson.
In models with extra dimensions, the Higgs boson can decay into a
neutrino pair~\cite{extra-dim} or can oscillate into invisible states
if the Higgs boson mixes with a graviscalar~\cite{graviscalar} which
is a scalar graviton and escapes in the extra dimension.
Models which introduce hidden scalar sectors which couple to the Higgs
sector can also cause invisible decays of the Higgs
boson~\cite{stealthy}.

In this paper, a search for invisibly decaying Higgs bosons
$(\ho\to\X\X)$\footnote{While motivated by the lightest neutralino of
  the MSSM, throughout this paper we use \X\ as a generic symbol for a
  neutral weakly interacting massive particle resulting from an
  invisible Higgs boson decay.} is presented using the data collected
at various centre-of-mass energies ($\sqrt{s}$) between 183 and
209~\gev\ by the OPAL detector at LEP, corresponding to an integrated
luminosity of \lumtot~\pb. The topology of events containing invisibly
decaying Higgs bosons produced through the process $\rmee\to\ho\Zo$ is
characterised by the decay products of the associated \Zo\ boson plus
large missing momentum and a visible mass (\MVIS) of the event
consistent with \mZ. Here it is also assumed that the decay width of
the Higgs boson is negligibly small. A search for invisibly
decaying Higgs bosons with large decay width is presented in
Ref.~\cite{wdw-inv}. The search presented here looks for a hadronic
decay of the \Zo\ boson in association with missing energy. To cover
other \Zo\ decay modes, the results from this search are combined with
the results of the decay-mode independent \ho\Zo\ search~\cite{DECind}
where the \Zo\ decays into \rmee\ or \mm. The results of the invisibly
decaying Higgs boson search at LEP~I~\cite{LEP1-inv} are also included to
enhance the sensitivities to lower Higgs boson masses.

The same analysis is applied to search for the production of nearly
invisibly decaying Higgs bosons: $\rmee\to\Zo\ho\to(\qq)(\X\XD)$ 
assuming a small mass difference $\dm=2$ and 4 \gev\ between \X\ and
\XD.  The standard neutralino searches~\cite{OPAL-neutralino} are
sensitive to cases with $\dm\geq 3$ \gev.  Similar searches for an
invisibly decaying Higgs boson have been carried out by the other LEP
experiments~\cite{inv-h-searches}.
\section {The OPAL Detector, Data and Event Simulation\label{apparatus}}
\subsection{The OPAL Detector}
The OPAL detector is described in detail in
Ref.~\cite{OPAL-detector}\@.  The central tracking system consisted of
a silicon micro-vertex detector, a vertex drift chamber, a jet chamber
and $z$-chambers.  In the range $\ACOST<0.73$, 159 points could
be measured in the jet chamber along each track\footnote{ A
right-handed coordinate system is adopted, where the $x$-axis points
to the centre of the LEP ring, and positive $z$ is along the electron
beam direction.  The angles $\theta$ and $\phi$ are the polar and
azimuthal angles, respectively.}. At least 20 points on a track could
be obtained over 96\% of the full solid angle.  The whole tracking
system was located inside a 0.435~T axial magnetic field.  A
lead-glass electromagnetic calorimeter (ECAL) providing acceptance
within $\ACOST<0.984$, together with pre-samplers and
time-of-flight scintillators, was located outside the magnet coil in
the barrel region and at the front end of each endcap.  The magnet
return yoke was instrumented for hadron calorimetry (HCAL), giving a
polar angle coverage of $\ACOST<0.99$, and was surrounded by
external muon chambers.  The forward detectors (FD) and
silicon-tungsten calorimeters (SW) located on both sides of the
interaction point measured the luminosity and complete the geometrical
acceptance down to 24~mrad in polar angle.  The small gap between the
endcap ECAL and FD was filled by an additional
electromagnetic calorimeter, called the gamma-catcher (GC), and a counter consisting of tile scintillators called the MIP plug.
\subsection{Data and event simulation}
The search is performed using the OPAL data collected at $\sqrts$
between 183 and 209 \gev\ with an integrated luminosity of \lumtot~\pb.
The integrated luminosities at each $\sqrts$ are listed in
Table~\ref{tab:zh-inv-lum}.
\begin{table}[htbp]
  \begin{center} 
    \begin{tabular}{|c|c|c|c|c|c|c|c|c|c|c|c|} 
      \hline
      \multicolumn{2}{|c|}{Nominal $\sqrt{s}$ (\gev) }&
      183&189&192&196&200&202&204&205&206&208\\ \hline
      \multicolumn{2}{|c|}{$\langle\sqrt{s}\rangle$ (\gev) }&
      \ecma&\ecmb&\ecmc&\ecmd&\ecme&\ecmf&\ecmg&\ecmh&\ecmi&\ecmj\\ \hline
      \multicolumn{2}{|c|}{ Lumi. (pb$^{-1}$)} &
      \luma&\lumb&\lumc&\lumd&\lume&\lumf&\lumg&\lumh&\lumi&\lumj \\ \hline
    \end{tabular}
    \caption{ \label{tab:zh-inv-lum} \sl Average centre-of-mass
      energies ($\langle\sqrt{s}\rangle$) and integrated luminosities collected
      at each nominal centre-of-mass energy after detector status cuts. The
      uncertainty on the luminosity measurement is $\it{0.5 \%}$.  }
  \end{center}
\end{table}

The signal detection efficiencies and expected number of background
events are estimated using a variety of Monte Carlo (MC) samples.
Signal samples of invisibly decaying and nearly invisibly decaying
Higgs boson processes are produced using the HZHA
generator~\cite{hzha}. The Higgs bosons are produced in association
with a \Zo\ boson, and then are forced to decay into a pair of
invisible particles. Samples of $\ho\to\X\X$ at each \sqrts\ are
produced in one \gev\ steps in the Higgs boson mass range
from 1 to 120 \gev\ with 2000 events per mass point.  The
$\ho\to\X\XD$ samples are generated with mass differences \dm\ of 2
and 4~GeV at 5 or 10 \gev\ Higgs boson mass intervals between 30 and
120 \gev, with 500 or 1000 events per point. The detection
efficiencies are determined at fixed values of the Higgs boson mass
using the above samples and then interpolated to arbitrary masses with
a spline fit.

The most important background processes are $\rmee\to\WW\to\lnu\qq$
and $\rmee\to\ZZ\to\nn\qq$. The first of these channels fakes a signal
when the lepton is within a jet or escapes detection along the beam
axis, and the second is an irreducible background for Higgs bosons
with masses in the vicinity of the \Zo\ boson mass. The radiative
multihadron process $\rmee\to\qq(\gamma)$ also contributes due to the
escape of photon into the beam pipe.

The background processes are simulated primarily by the following
event generators. For two-fermion (2f) final states, events are
generated by PYTHIA~\cite{pythia} and KK2f~\cite{KK2f}
($\qq(\gamma$)), BHWIDE~\cite{bhwide} and TEEGG~\cite{teegg}
(\rmee$(\gamma)$), and KORALZ~\cite{koralz} and KK2f (\mm$(\gamma)$
and \tautau$(\gamma)$), for four-fermion (4f) final states, by
grc4f~\cite{grc4f} (4f processes with final states of
$e^+e^-f\bar{f}$) and KORALW~\cite{koralw} (4f processes except final
states with $e^+e^-f\bar{f}$), and for so-called two-photon processes
where the initial-state electron and positron radiate photons which
interact to produce additional final state fermions, by
PHOJET~\cite{phojet}, PYTHIA and Vermaseren~\cite{vermaseren}
(hadronic and leptonic two-photon processes; $\rmee\qq$ and
\rmee\ellell). The generated partons are hadronised using
JETSET~\cite{pythia} with parameters described in
Ref.~\cite{opaltune}. The resulting particles are processed through a
full simulation~\cite{gopal} of the OPAL detector.
\section {Selection criteria \label{analysis}}
The search criteria are optimised at each \sqrts\ using the MC samples
with 10 mass points just below the kinematic limit for the invisibly
decaying Higgs, $\ho\to\X\X$.  For the $\ho\to\X\XD$ final state, the
decay products of the \Zo\ may be accompanied by a soft jet with small
visible mass and energy, aligned in the direction of the missing
momentum.  Since the two event topologies are very similar, the
selection criteria for the $\ho\to\X\X$ are also applied to the
$\ho\to\X\XD$ final states.  The analysis begins with a preselection
to ensure data quality, followed by a combination of cut-based and
likelihood-based analysis.

Experimental variables are calculated using the four-momenta of
charged particle tracks, and ECAL and HCAL clusters. The clusters
associated with tracks are also used in the energy and momentum
calculations, after subtracting the momenta of tracks from the energy
observed in the calorimeters to reduce double counting of
energy~\cite{mt300-lepton-jet}.
\subsection {Preselection \label{sec:inv_pre}}
The following requirements are applied to reduce beam-related
background as well as two-photon events:
\begin{itemize}
\item[(P1)] The event must not contain any charged particle track or
  ECAL cluster with reconstructed energy greater than $1.3 \times\EBEAM$, where
  \EBEAM\ is the beam energy.
\item[(P2)] $\EVISCT/\EVIS < 0.2$, where \EVIS\ is
  the total visible energy and \EVISCT\ is the visible energy in the
  region defined by $\ACOST > 0.9$.
\item[(P3)] $\NGCH/\NCH>0.2$, where \NGCH\ and \NCH\ are the number of
  good charged particle tracks defined as in Ref.~\cite{gchtr} and
  total number of tracks, respectively.
\item[(P4)] $\MVIS>3~\gev$, where
  \MVIS\ is the invariant mass of the event.
\item[(P5)] $\pt >1.8~\gev$, where \pt\ is the
  magnitude of the vector sum of the transverse momenta of the
  reconstructed objects in the event with respect to the beam
  direction.
\item[(P6)] Forward energy veto: events are rejected if there is more
  than 2/2/5~GeV deposited in either side of the forward detectors,
  SW/FD/GC respectively, or if there is any significant activity in
  the MIP plug.  This forward energy veto is introduced to ensure that
  the data sample consists of well-measured events. The efficiency
  loss due to vetoes on random detector occupancy has been studied
  with a sample consisting of random triggers, and was found to be
  between $2.2\%$ and $4.1\%$, depending on \sqrts.  The signal
  detection efficiencies and the numbers of expected background events
  are corrected for such losses.
\item[(P7)] $\njets>1$, where \njets\ is the
  number of jets reconstructed with the Durham algorithm~\cite{durham}
  with a jet resolution parameter \ycut\ = 0.005. This reduces
  monojet-like background caused primarily by beam-gas and beam-wall
  interactions.
\item[(P8)] $\ACMIS<0.95$, where \TMIS\ is the polar angle of the
  missing momentum of the event.  The $\qq(\gamma)$ background is
  reduced by this requirement.
\item[(P9)] $\MMISSQ>0~\gev^2$, where
  \MMISSQ\ is missing mass squared and is calculated with
  the visible mass scaled to the \Zo-mass, {\it i.e.}
  $\MMISSQ = s-2 \sqrt{s}
  \frac{\mZ}{\MVIS}\EVIS+\mZSQ$. This formula is applied
  to avoid a negative \MMISSQ.
\end{itemize}

The number of data events remaining after these cuts and those
expected from SM background processes are summarised in the first row
of Table~\ref{tab:cut-selection}.
\subsection{Main selection criteria \label{sec:inv_qq}}
\begin{figure}[htbp]
  \centerline{
    \epsfig{file=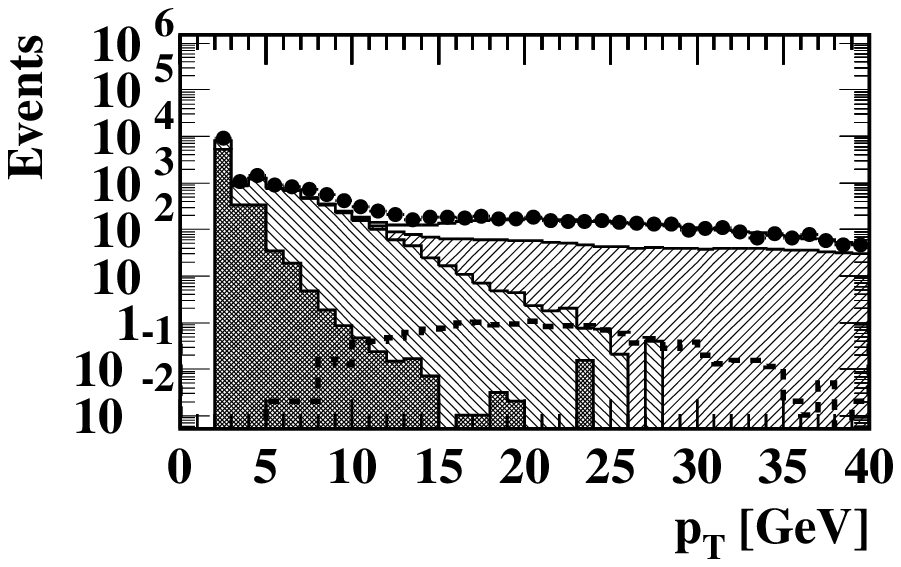,height=0.4\textheight} }
  \centerline{
    \epsfig{file=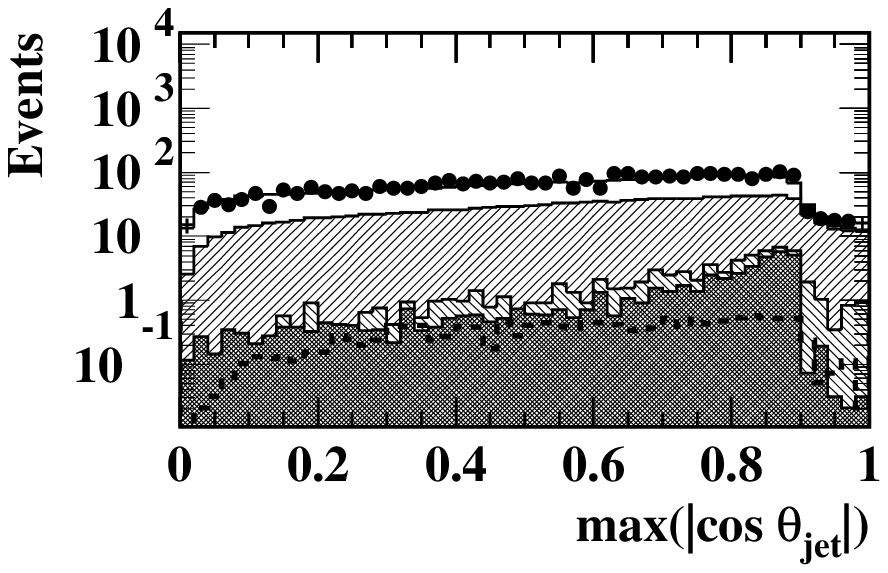,height=0.4\textheight} }
    \caption[]{\label{fig:cut-selection}\sl The distributions of
      \pt\ and \ACJET\ at $\sqrt{s}$=206~\gev\ before
      applying the sequential cuts on the variables: data (dots with
      error bar) are shown together with the predicted contributions
      from the background processes; $\rmee\to\ellell$
      (cross-hatched), two-photon processes (negative slope hatched),
      four-fermion processes (positive slope hatched), and
      $\qq(\gamma)$ (open).  The background distributions have been
      normalised to \lumi~\pb.  The distribution of simulated signals
      for the process $\Zo\ho\to(\qq)(\X\X)$ for $\mh$ =
      \mhsampA~\gev\ are also shown with dashed line.  The signal is
      normalised using the SM Higgs production cross-section and 100\%
      production rate for the process $\Zo\ho\to(\qq)(\X\X)$.}
\end{figure}

The main selection consists of a cut-based analysis followed by a
likelihood-based analysis using the same technique as described in
Ref.~\cite{172-sm-Higgs}. After the preselection (P1-P9) the
following cuts are applied in sequence:
\begin{itemize}
\item[(B1)] $\NGCH>4$.
\item[(B2)] $\pt>6~\gev$.
\item[(B3)] $\MACJET<0.95$, where \TJET\ is the polar angle of the
  jet axis after the event is forced into two jets with the Durham
  algorithm. This requirement leaves events containing well measured jets.
\item[(B4)] The number of isolated charged leptons identified as
  in Ref~\cite{172-sm-Higgs} is required to be zero to reduce the
  background contribution from semi-leptonic \WW\ and \ZZ\ events.  
\item[(B5)] $120~\gev>\MVIS>50~\gev$.
\end{itemize}
The distributions of \pt\ and
\MACJET, just before applying the respective cuts,
are shown in Figure~\ref{fig:cut-selection}. The numbers of selected
events, the expected background and the signal efficiencies, after
each cut, are shown in Table~\ref{tab:cut-selection}.
\begin{table}[htbp]
  \begin{center}
    \begin{tabular}{|cc||r||r||r|r||c|}\hline
      \multicolumn{2}{|c||}{Cut} &Data&
      \multicolumn{3}{|c||}{Background}&Efficiency (\%) \\ \cline{4-6}
      & & & \multicolumn{1}{|c||}{Total} & 
      \multicolumn{1}{|c|}{2f} & \multicolumn{1}{|c||}{4f} 
      & $\mh = \mhsampA~\gev$   \\ \hline
      \multicolumn{2}{|c||}{Pre}& 101653& 86767
& 48277& 8299&  67.6\\\hline
      \multicolumn{2}{|c||}{B1}&  40031&  34158
& 14253& 7513&  67.6 \\
      \multicolumn{2}{|c||}{B2}&  16895&  17037
& 10391& 6503&  65.8 \\
      \multicolumn{2}{|c||}{B3}&  16694&  16882
& 10372& 6379&  65.3 \\
      \multicolumn{2}{|c||}{B4}&  11476&  11654
&  8855& 2695&  56.1 \\
      \multicolumn{2}{|c||}{B5}&   1045&   1069
&  523&  532&   55.5\\\hline\hline
      LH&\jj & 194&$205.7\pm 1.8 \pm 1.3 $
& 46.1&157.1& 23.7\\
      &\jjj & 278&$279.4\pm 1.9 \pm 1.5 $
& 44.1&233.5& 23.4\\
      \hline
    \end{tabular}
    \caption[]{\label{tab:cut-selection}\sl Cut flow table at
      $\sqrt{s}=$183--209~GeV. Each row shows the number of events
      after each cut of the selection (described in the text) for the
      data and the expected background. The backgrounds from
      two-fermion and four-fermion processes are shown separately. The
      contributions from two-photon processes are not shown
      individually but included in the total background. The
      background estimates are normalised to luminosity at each energy
      and summed.  The first quoted error on background estimates is
      statistics and the second systematic.  The last column shows the
      luminosity-weighted average of selection efficiencies for the
      $\Zo\ho\to(\qq)(\X\X)$ final state with $\mh$ = \mhsampA~\gev.
      The last two rows show the final numbers of selected events,
      expected background and the efficiency after the likelihood
      analysis (LH) in each category.  The efficiency in a category is
      the fraction of signal Monte Carlo events which pass the
      selection requirements.  The background numbers and signal
      efficiencies include the occupancy correction determined at each
      \sqrts\ due to the forward energy veto.}
  \end{center}
\end{table}

\begin{figure}[htbp]
  \vspace{-1.6cm} \centerline{
    \includegraphics[width=0.45\linewidth,clip]{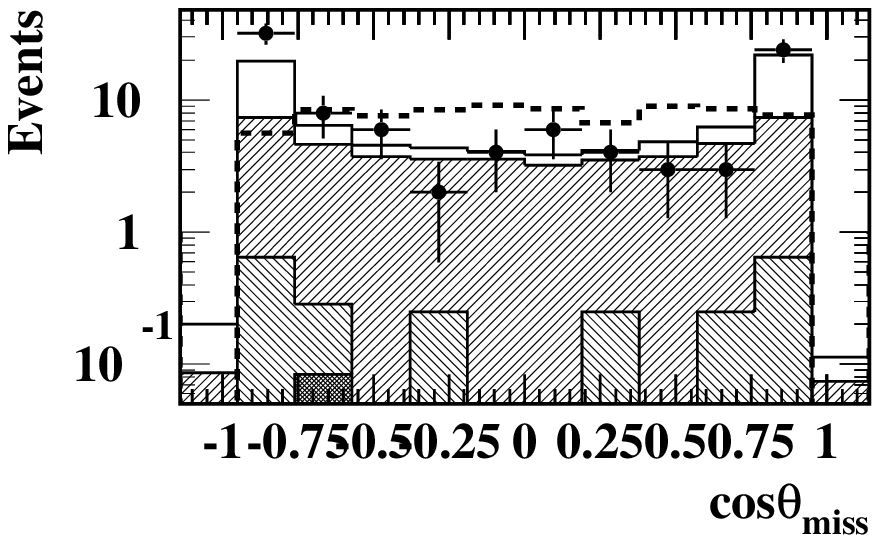}
    \includegraphics[width=0.45\linewidth,clip]{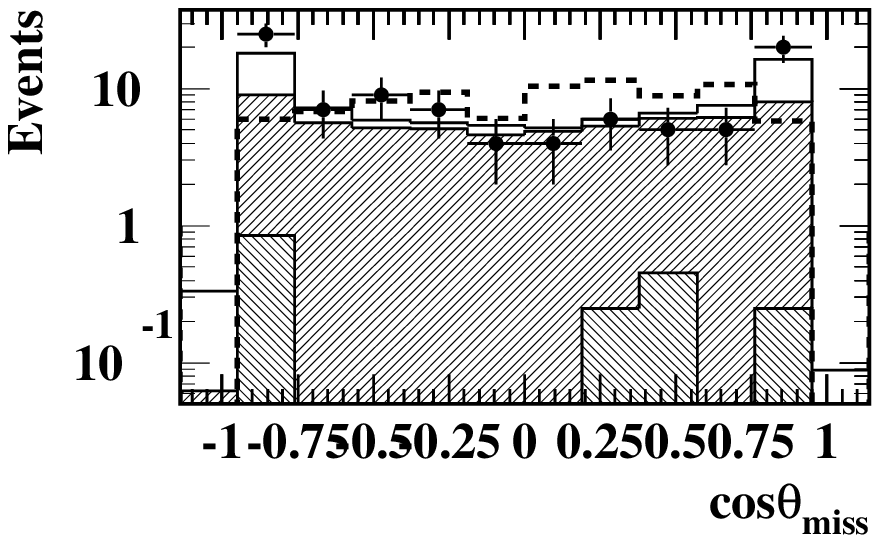}
  }
  \vspace{-1.6cm} \centerline{
    \includegraphics[width=0.45\linewidth,clip]{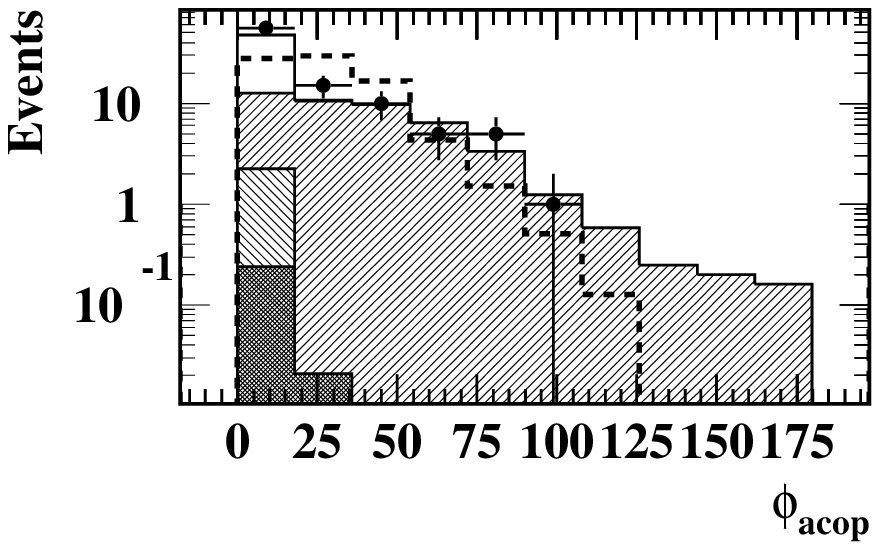}
    \includegraphics[width=0.45\linewidth,clip]{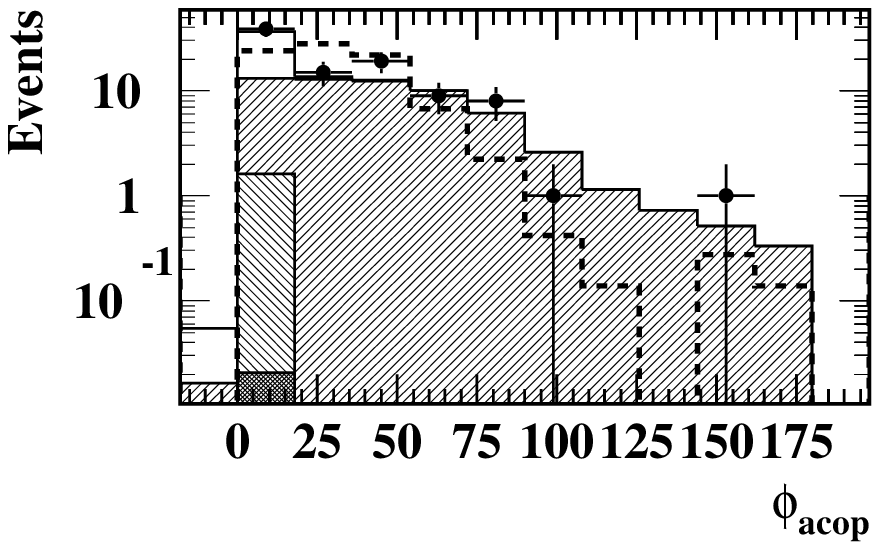}
  }
  \vspace{-1.6cm} \centerline{
    \includegraphics[width=0.45\linewidth,clip]{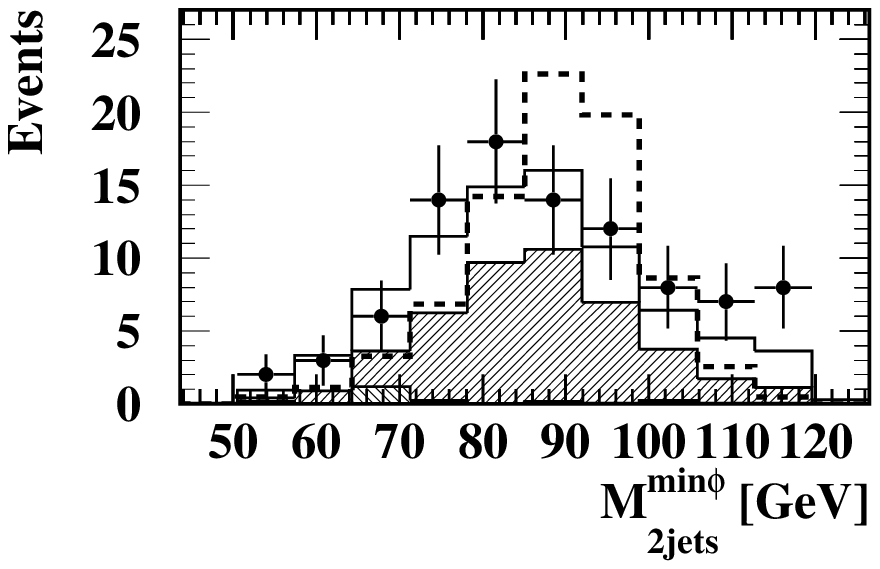}
    \includegraphics[width=0.45\linewidth,clip]{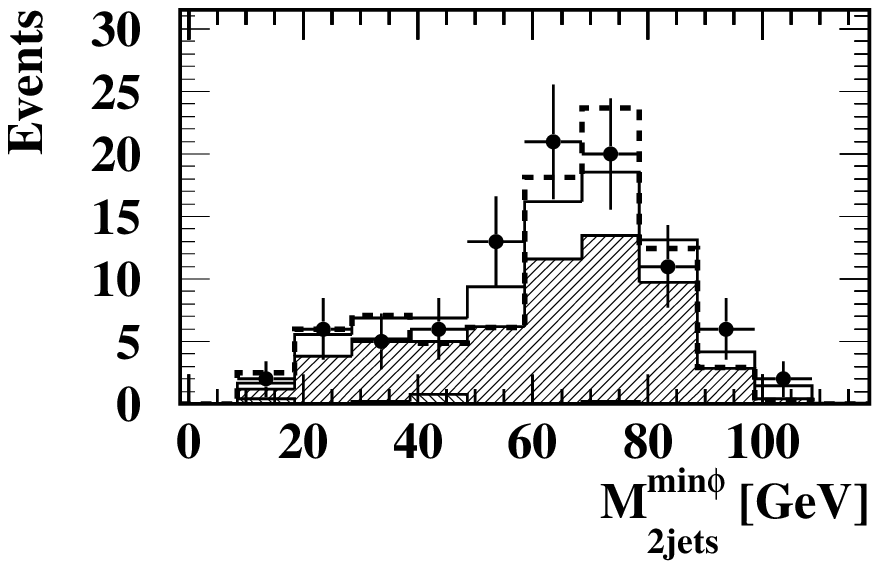}
  }
  \vspace{-1.6cm} \centerline{
    \includegraphics[width=0.45\linewidth,clip]{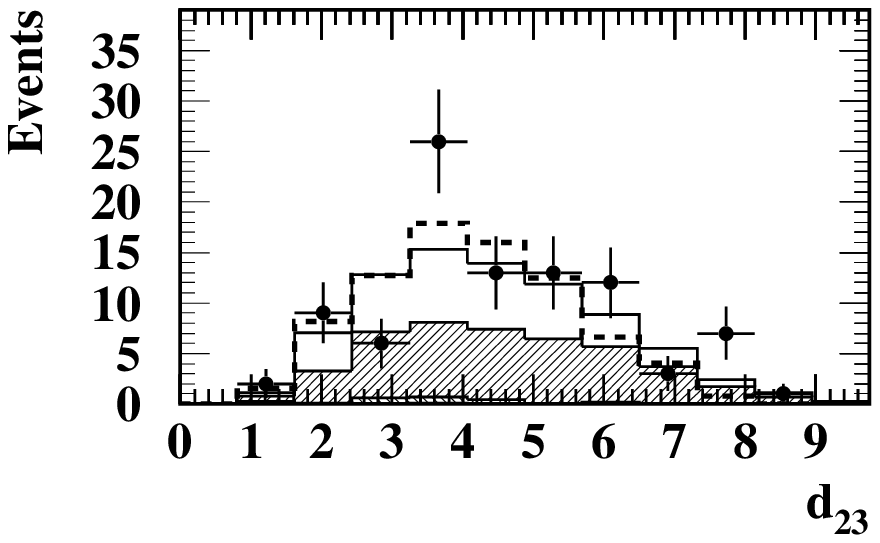}
    \includegraphics[width=0.45\linewidth,clip]{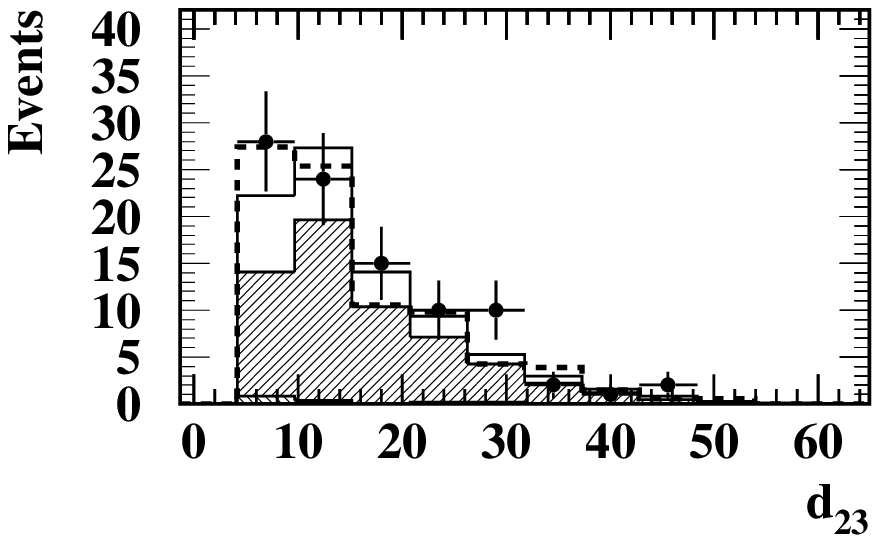}
  }
  \vspace{-1.6cm} \centerline{
    \includegraphics[width=0.45\linewidth,clip]{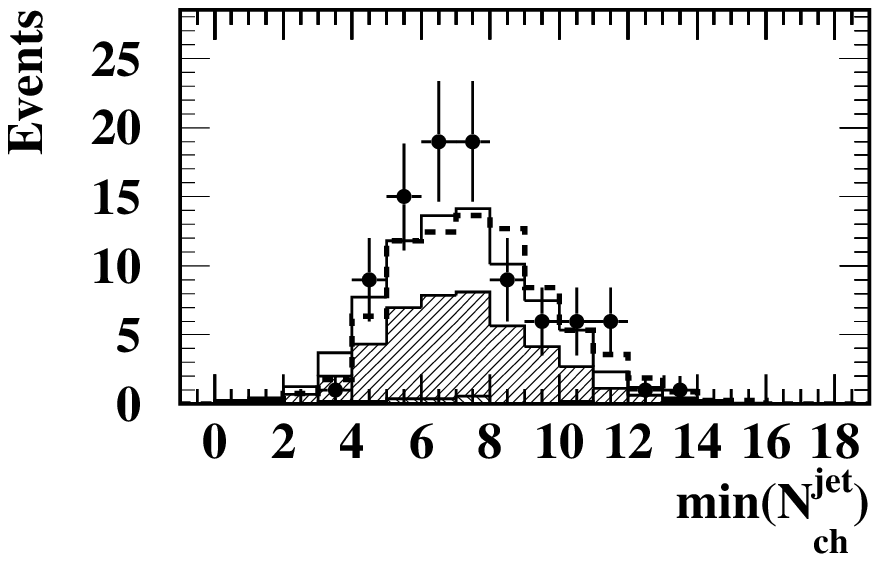}
    \includegraphics[width=0.45\linewidth,clip]{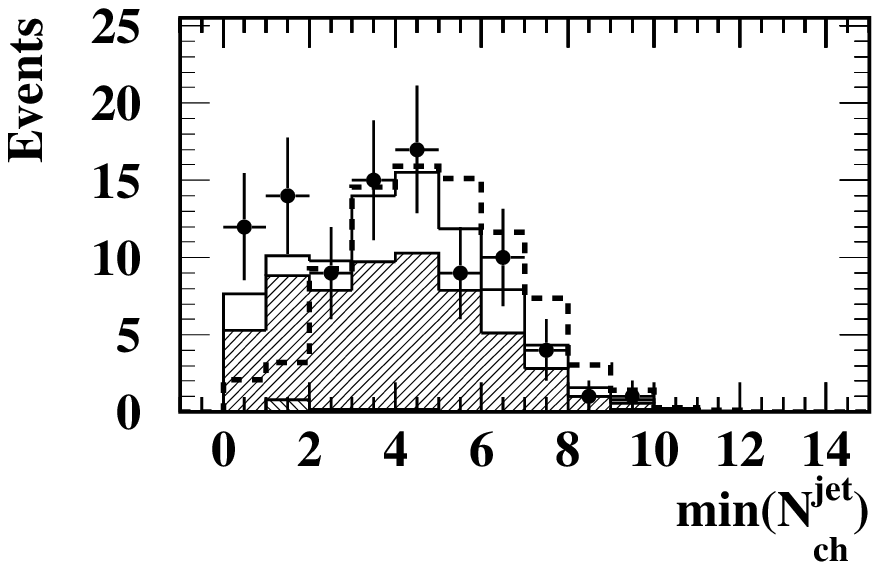}
  }
  \caption[]{\label{fig:lh-inputs}\sl The distribution of likelihood
    input variables: \CMIS, \acop,
    $M_{\mathrm{2jets}}^{\mathrm{min\phi}}$, \DTT\ and
    $\mathrm{min}(N^{\mathrm{jet}}_{\mathrm{ch}})$, at
    $\sqrts=206$~\gev\ for the \jj\ category (left) and the
    \jjj\ category (right). The background sources are shaded as in
    Figure~\ref{fig:cut-selection}. The distributions of the signal
    for simulated invisibly decaying Higgs bosons with $\mh$ =
    \mhsampA~\gev\ are shown as dashed lines. The signal histograms
    are normalised to the number of events of the expected
    background. The first and last bins in each histogram include
    underflow and overflow, respectively.}
\end{figure}

\begin{figure}[htbp]
  \centerline{
    \includegraphics[width=0.52\linewidth,clip]{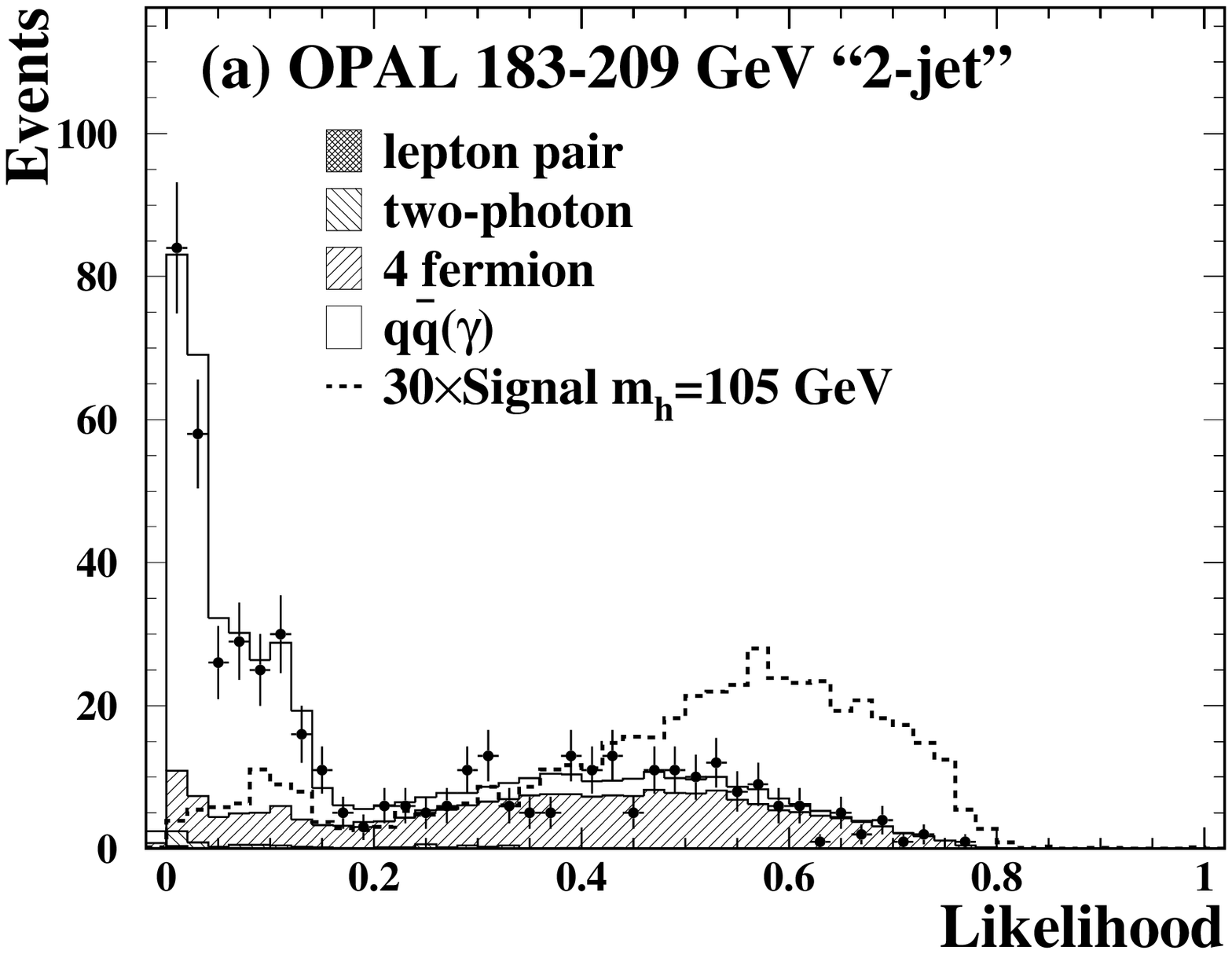}
    \includegraphics[width=0.52\linewidth,clip]{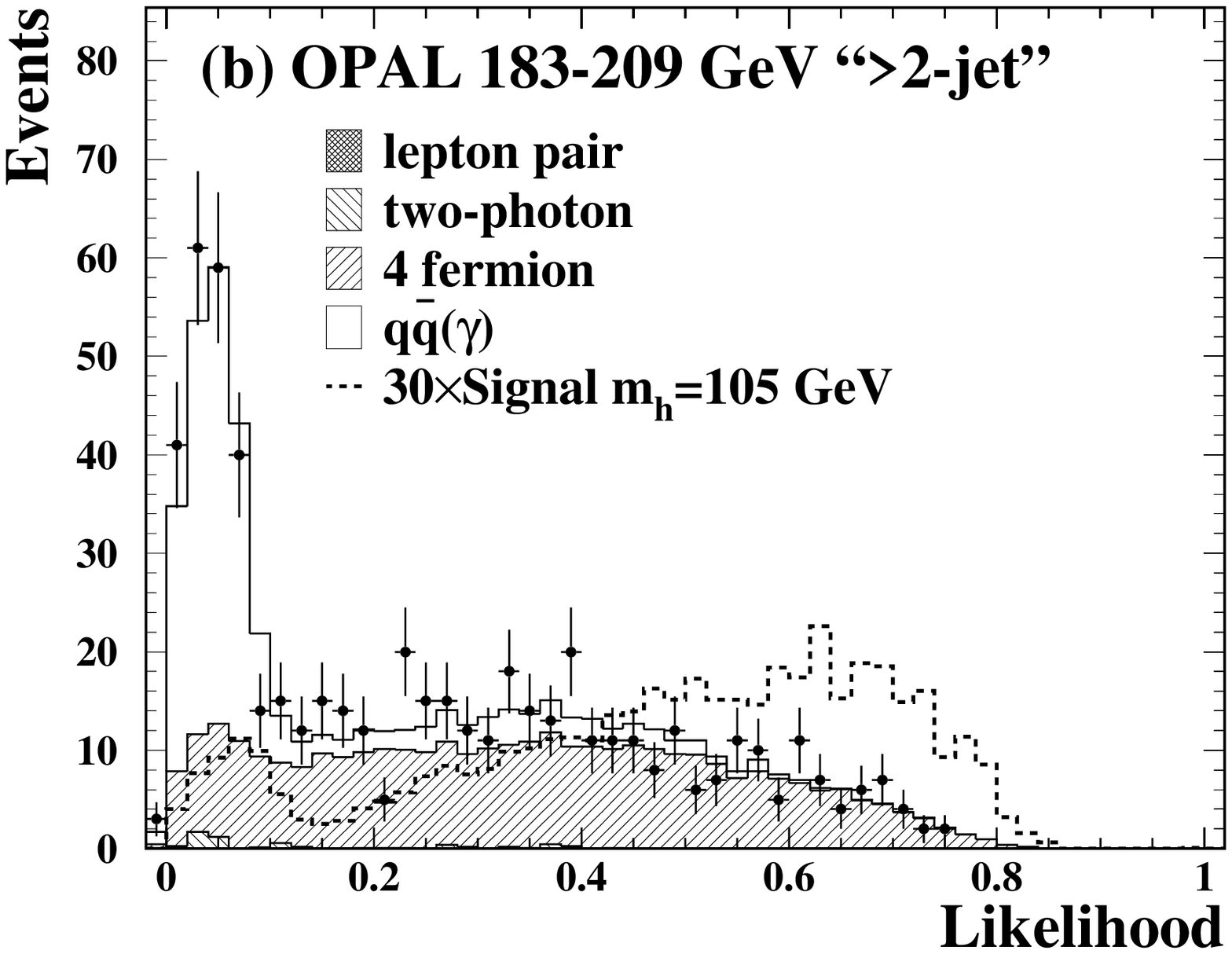}
  }
  \caption[]{\label{fig:likelihood}\sl The distribution of likelihood
    output for \sqrts= 183 -- 209~\gev\ for (a) the \jj\ category and
    (b) the \jjj\ category. The background sources are shaded as in
    Figure~\ref{fig:cut-selection}. The distributions of the signal
    for simulated invisibly decaying Higgs bosons with $\mh$ =
    \mhsampA~\gev\ are also shown.  The signal histograms are
    normalised using 30 times the production cross-section of the SM
    Higgs boson and 100\% production rate for the process
    $\Zo\ho\to(\qq)(\X\X)$.  }
\end{figure}

After applying the above cuts, the selected sample is divided into two
categories, namely events with two jets (``\jj '') and with more than
two jets (``\jjj''), where the number of jets is defined by the Durham
algorithm with \ycut\ = 0.005. A likelihood analysis (LH) is built up
for each category separately, with the same technique as described in
Ref.~\cite{172-sm-Higgs} using the input variables:
\begin{itemize}
\item{\CMIS}
\item{the acoplanarity angle when the event is forced into two jets, \acop.}
\item{the invariant mass of the two jets with the smallest opening
  angle, $M_{\mathrm{2jets}}^{\mathrm{min\phi}}$.}
\item{\DTT, which is defined as $\EVISSQ\times\YTT$, where \YTT\ is
  the jet resolution parameter at the transition point from two to
  three jets in the jet reconstruction.}
\item{$\mathrm{min}(N^{\mathrm{jet}}_{\mathrm{ch}})$, which is
  the smallest charged multiplicity of any jet in the event.}
\end{itemize}
The distributions of input variables for the expected background are
different between the two categories as shown in
Figure~\ref{fig:lh-inputs}, due to the different contribution from
background sources. The resulting likelihood distribution for each
category is shown in Figure~\ref{fig:likelihood}. The remaining
background in the signal-like region is dominated by semi-leptonic 4f
events.

The properties of 4f background events are similar to the signal,
thus broadening the likelihood peak for the signal. The final results are
obtained by requiring the likelihood to be larger than~0.2. The
numbers of observed and expected events are summarised in
Table~\ref{tab:cut-selection} and Table~\ref{tab:zh-inv-cutflow}.  The
efficiency in a category is defined as the ratio of the number of
selected events in that category to the total number of produced
events; the sum of efficiencies in the two categories provides the
total efficiency at a given mass point. The efficiencies for the
$\ho\to\X\XD$ processes are relatively lower than those for the
$\ho\to\X\X$, as shown in Table~\ref{tab:zh-inv-eff}.    
\begin{table}[htbp]
  \begin{center} 
    \begin{tabular}{|c|c|l|l|l|l|l|l|l|l|l|l|} 
      \hline
      \multicolumn{2}{|c|}{}&
      \multicolumn{10}{|c|}{$\sqrt{s}$ (GeV)}\\
      \cline{3-12}
      \multicolumn{2}{|c|}{}&183&189&192&196&200
      &202&204&205&206&208\\ \hline
      &Data
      &17&52& 7&19&20&12&1&22&43& 1\\ \cline{2-12}
      \jj &Background
      &19.0&54.6& 9.0&22.4&23.4&12.2& 2.1&22.1&38.5& 2.4\\ \cline{2-12}
      &Eff.(\%)
      &17.7&20.0&21.1&22.5&27.2&26.7&28.5&27.6&27.2&29.2\\ \hline
      &Data 
      &15&78&18&31&41&12& 2&32&47& 2\\ \cline{2-12}
      \jjj &Background
      &30.6&76.1&13.1&31.7&30.8&15.4& 2.6&28.5&47.6& 3.1\\ \cline{2-12}
      &Eff.(\%)
      &17.5&20.9&20.7&24.3&25.0&25.1&24.9&23.6&26.1&23.6\\ \hline
    \end{tabular}
    \caption{ \label{tab:zh-inv-cutflow} \sl Number of candidate
      events and expected background for each category at each \sqrts,
      together with signal efficiencies for $\mh=\mhsampA~\gev$.  }
  \end{center}
\end{table}
\begin{table}[htbp]
  \begin{center} 
    \begin{tabular}{|c|c|c|l|l|l|l|l|l|l|l|l|l|l|} 
      \hline
      Cat.&Decay&\dm&\multicolumn{11}{|c|}{Efficiencies (\%) at Higgs
      Mass(GeV)}\\ \cline{4-14} 
      &Mode&(GeV)&10&20&30&40&50&60&70&80&90&100&105\\ \hline 
      \jj &\X\X&---
      &13&16&20&22&26&27&27&27&27&25&24\\ \cline{2-14}
      &\X\X$\gamma$&2
      &---&---&11&14&18&22&24&23&25&23&24\\ \cline{2-14}
      &\X\X$\gamma$&4
      &---&---& 8&11&14&17&19&20&20&20&21\\ \cline{2-14}
      &\X\X\Zs&2
      &---&---& 9&13&16&19&22&22&23&21&22\\ \cline{2-14}
      &\X\X\Zs&4
      &---&---& 9&13&16&18&21&22&22&21&22\\ \hline\hline
      \jjj &\X\X&---
      &12&14&18&20&23&24&24&25&25&23&23\\ \cline{2-14}
      &\X\X$\gamma$&2
      &---&---&10&14&17&19&22&23&23&22&23\\ \cline{2-14}
      &\X\X$\gamma$&4
      &---&---&10&14&17&19&22&23&24&23&24\\ \cline{2-14}
      &\X\X\Zs&2
      &---&---&12&15&18&20&23&24&24&23&24\\ \cline{2-14}
      &\X\X\Zs&4
      &---&---&12&16&19&21&23&24&25&24&25\\ \hline
    \end{tabular}
    \caption{\label{tab:zh-inv-eff} \sl Luminosity-weighted averages
      of efficiency for \X\X, \X\X$\gamma$ and \X\X\Zs\ channels. No
      Monte Carlo samples are available in the \X\X$\gamma$ and
      \X\X\Zs\ channels for Higgs masses less than 30 \gev.  }
  \end{center}
\end{table}

The systematic errors on signal efficiencies and the numbers of
expected background events are estimated using the following
procedures.  The uncertainty corresponding to the modelling of each
selection variable is determined by comparing the mean values of the
distribution of that variable between data and SM background MC
samples at $\sqrt{s}=\mZ$ after applying the preselection.
Efficiencies and numbers of expected background events are estimated
again, shifting each variable separately by its uncertainty.  Relative
changes to the original values of efficiencies and numbers of expected
background events are taken as systematic errors for that
variable. The systematic errors for the LH selection are estimated in
a similar way.  The total systematic errors due to the modelling of
the selection variables including those entering the LH selection are
calculated by summing the errors in quadrature for each category at
each \sqrts\ individually.
The evaluated errors are summarised in Table~\ref{tab:syst-err}.
The statistical errors due to the finite size of the MC samples and
the uncertainty on the luminosity measurement are also estimated. The
total systematic error ranges from $3.5\%$ to $17.4\%$ for the
signal, and from $1.9\%$ to $4.4\%$ for the background.
\begin{table}[htbp]
  \begin{center} 
    \begin{tabular}{|c|c|c|c|c|c|c|c|} 
      \hline
      \multicolumn{2}{|c|}{Category}&\multicolumn{3}{|c|}{\jj}&
      \multicolumn{3}{|c|}{\jjj}\\\hline
      \multicolumn{2}{|c|}{Decay Mode}&Inv.&\multicolumn{2}{|c|}{Nearly Inv.}&
      Inv.&\multicolumn{2}{|c|}{Nearly Inv.}\\\hline
      \multicolumn{2}{|c|}{\dm\ (\gev)}&---&$2$&$4$&---&$2$&$4$ \\\hline
      Selection variable & Signal& 0.2-3.9\%& 0.0-5.0\%& 0.0-11.1\%& 0.3-4.1\%&
      0.4-4.7\%& 0.6-10.3\% \\\cline{2-8} 
       & BKG& \multicolumn{3}{|c|}{0.7-2.2\%}& \multicolumn{3}{|c|}{1.0-1.9\%} \\\hline
      MC statistics&Signal& 3.3-7.3\%& 5.2-10.9\%& 6.5-15.4\%& 3.6-8.8\%&
      5.0-9.5\%& 4.8-11.0\% \\\cline{2-8}
       & BKG& \multicolumn{3}{|c|}{1.9-3.8\%}& \multicolumn{3}{|c|}{1.4-2.7\%} \\\hline
      \multicolumn{2}{|c|}{Luminosity}& \multicolumn{6}{|c|}{$0.5\%$} \\\hline
      \hline
      Total&Signal& 3.5-7.8\%& 5.3-10.7\%& 5.8-17.4\%& 3.7-9.5\%&
      5.1-7.8\%& 4.8-11.4\% \\\cline{2-8}
      & BKG& \multicolumn{3}{|c|}{2.2-4.4\%}& \multicolumn{3}{|c|}{1.9-3.4\%} \\\hline
    \end{tabular}
    \caption{ \label{tab:syst-err} \sl Ranges of estimated relative
      systematic errors (in \%) for all \sqrts. Errors on the signal
      efficiency are estimated at each MC mass point at each \sqrts\
      and those for background at each \sqrts. The total systematic
      error at each \sqrts\ is calculated by summing the individual
      errors in quadrature.}
  \end{center}
\end{table}
\section{Results\label{sec:inv_res}}
\begin{figure}[htbp]
  \centerline{
    \includegraphics[width=0.52\linewidth,clip]{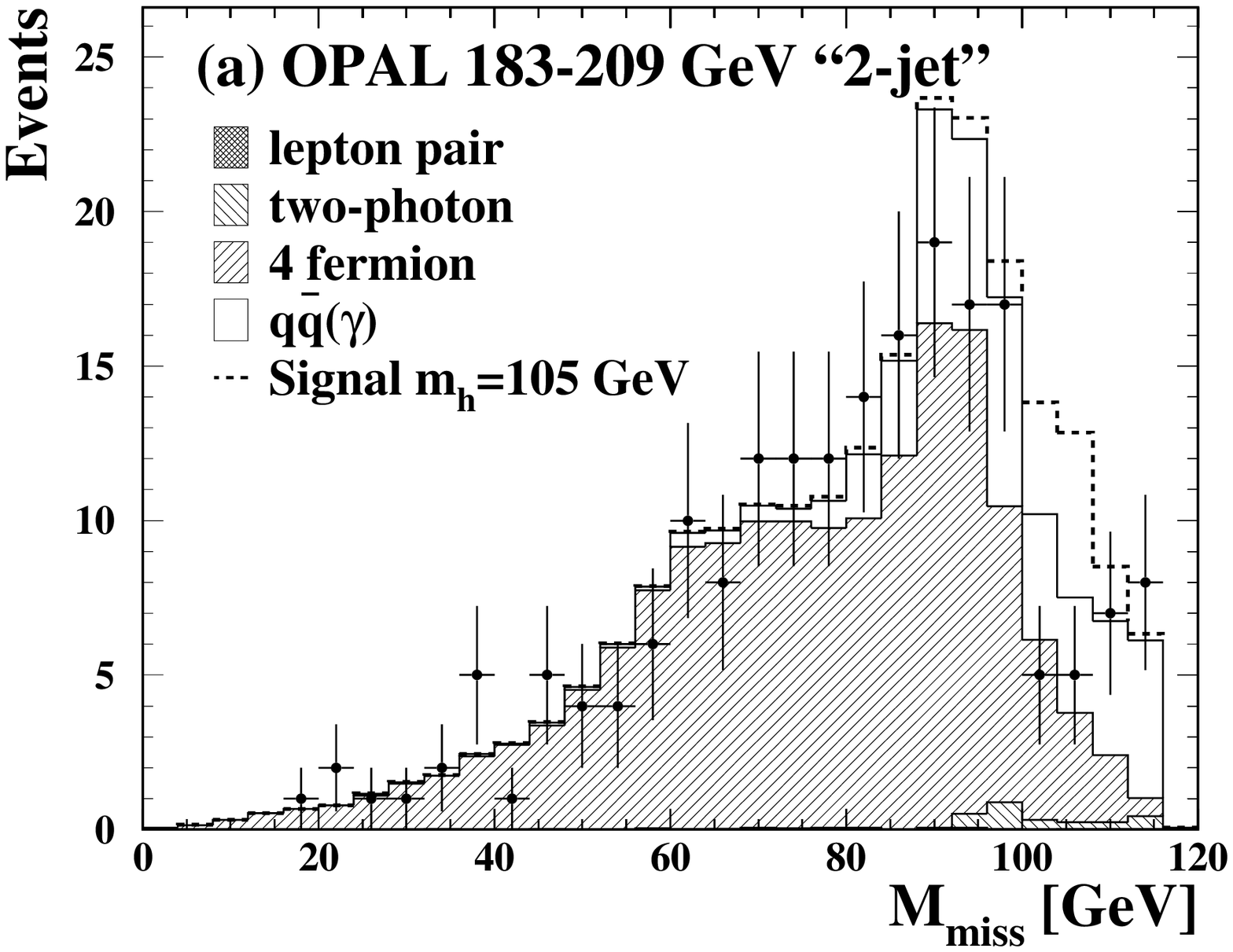}
    \includegraphics[width=0.52\linewidth,clip]{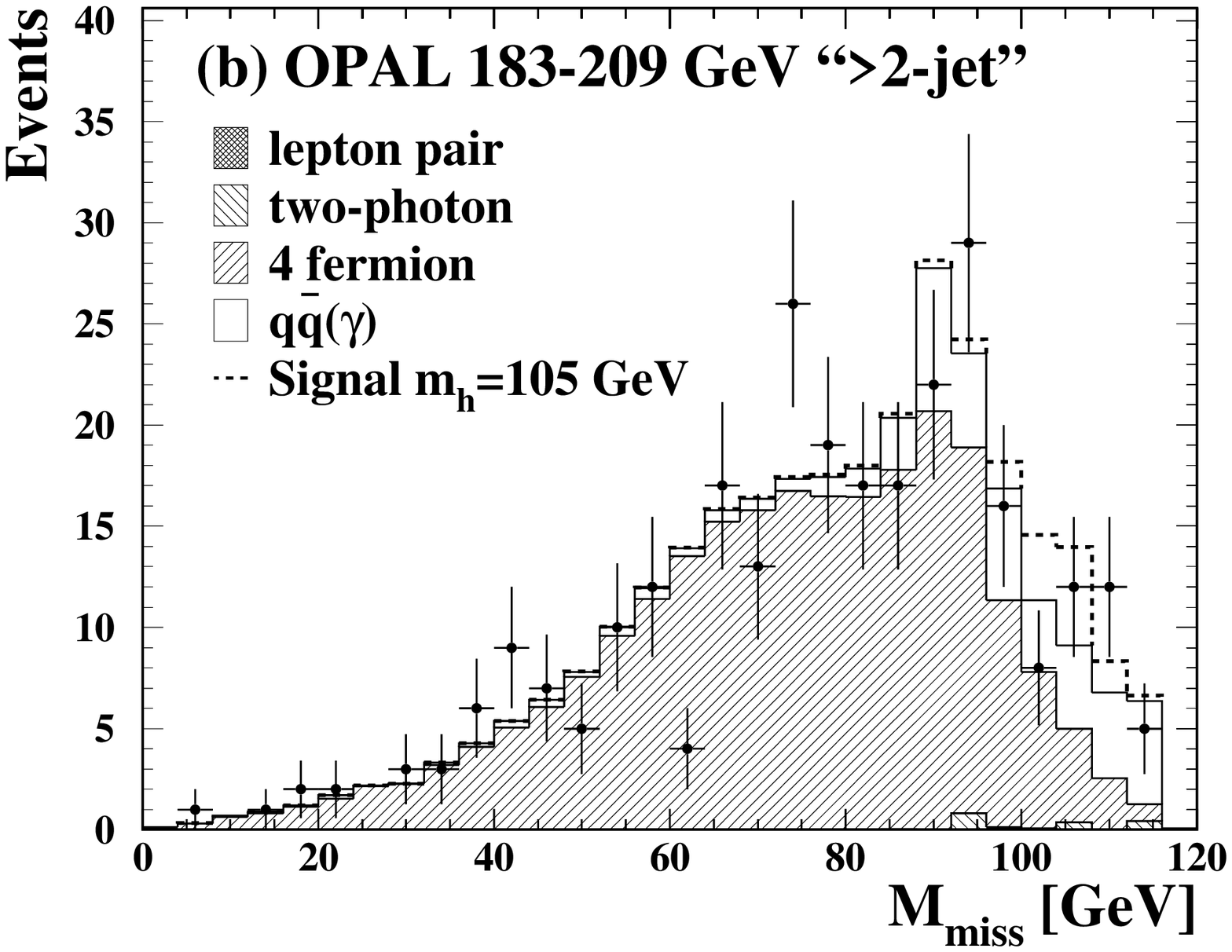}
  }
  \caption[]{\label{fig:mmis}\sl The distribution of missing mass for
    each category for all LEP2 data combined: (a) the \jj\ category
    and (b) the \jjj\ category.  The background sources are shaded as
    in Figure~\ref{fig:cut-selection}.  The distributions of the
    signal for simulated invisibly decaying Higgs bosons with $\mh$ =
    \mhsampA~\gev\ are shown on top of the background
    distribution. The signal histograms are normalised using the
    production cross-section of the SM Higgs boson and 100\%
    production rate for the process $\Zo\ho\to(\qq)(\X\X)$.  }
\end{figure}

Figure~\ref{fig:mmis} shows the missing mass distribution for the
selected candidate events together with the expected background and an
expected signal of $\mh=\mhsampA~\gev$ for the two categories, for all
\sqrts\ combined.  No significant excess above the expected SM
background is observed in either category. The main background comes
from four-fermion processes in both categories. The broad peak around
70~GeV in the four-fermion histograms is due to the $\rmee\to\WW$
process and the peak around 90~GeV is due to the $\rmee\to\ZZ$
process. Final efficiencies are summarised in
Table~\ref{tab:zh-inv-cutflow}.

Limits are calculated using the likelihood ratio method described in
Ref.~\cite{tj-method}. The \MMIS\ information is used as a
discriminator in the calculation. Systematic errors on the background
and signal estimate are taken into account.
\subsection{Limits on the production of invisibly 
  decaying Higgs bosons\label{sec:inv_rinv}}
\begin{figure}[tbp]
  \centerline{
    \epsfig{file=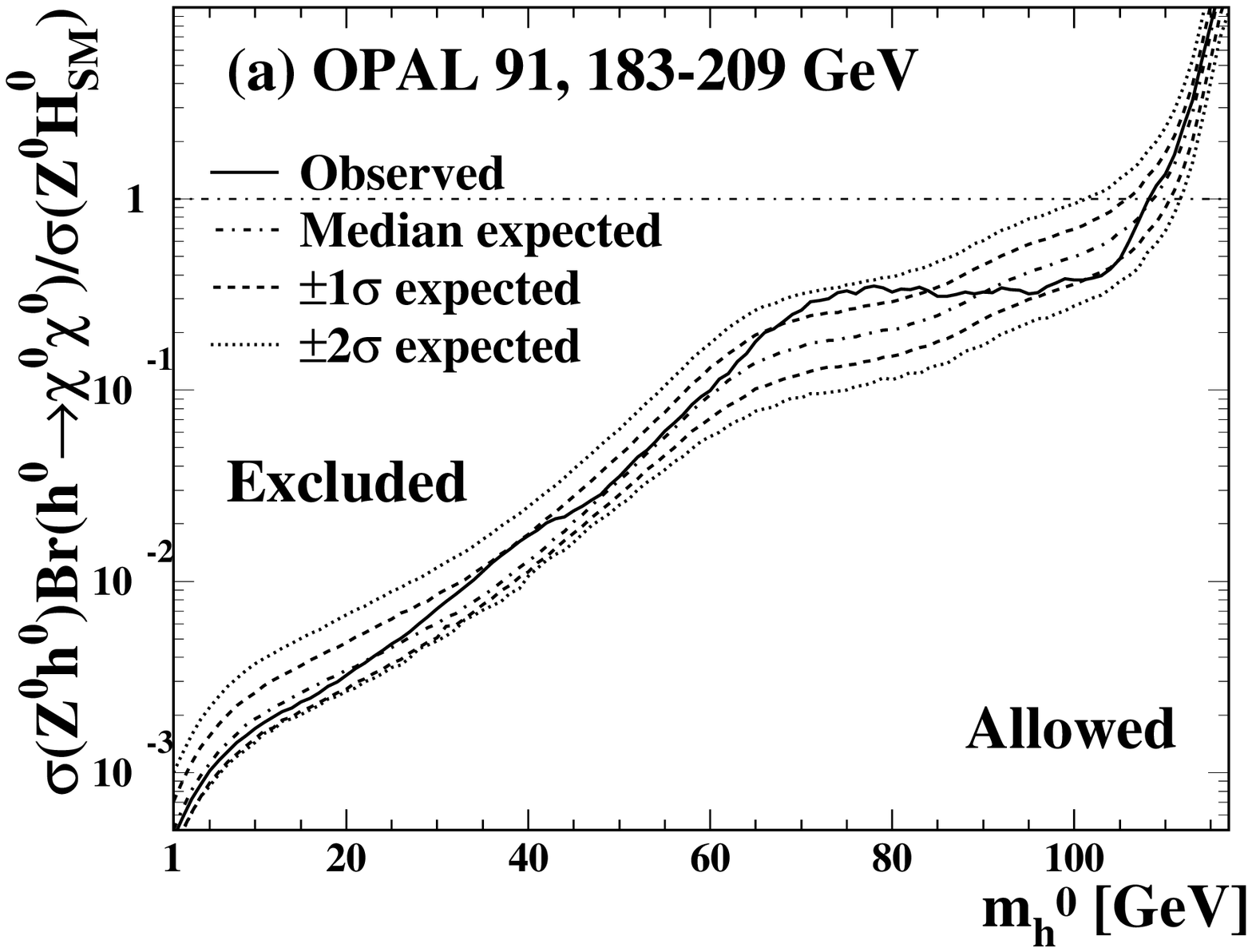,height=0.4\textheight} }
  \centerline{
    \epsfig{file=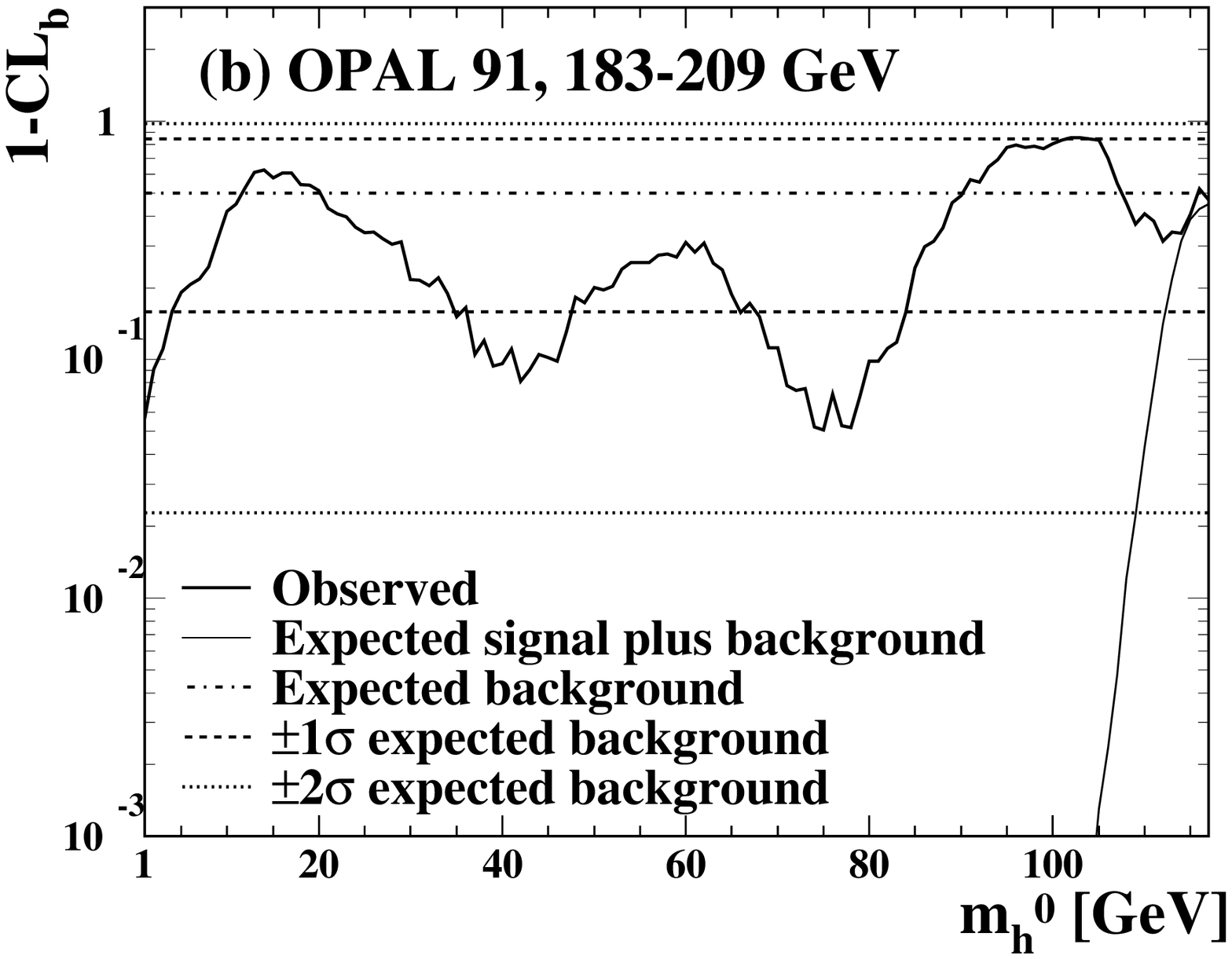,height=0.4\textheight} }
  \caption[]{\label{fig:zh_inv_linv}\sl (a) Observed and expected
    limits on the relative production rate for
    $\rmee\to\Zo\ho\to\Zo\X\X$ (invisible decay) to the SM Higgs
    production rate at 95\% CL as a function of the test mass \mh,
    assuming $\BR(\ho\to\X\X)=100\%$.  The solid curves show the
    observed limits and the dot-dashed curves the median expected
    limits. The dashed and dotted curves show $1\sigma$ and $2\sigma$
    bands of expected limits.  (b) The background confidence \clb\ as
    a function of \mh. The thick solid curve shows the observed \clb\
    and the thin solid curve the expectation in the signal plus
    background hypothesis.  The dot-dashed, dashed and
    dotted lines show median \clb, and the $1\sigma$ and $2\sigma$
    bands expected for the background only hypothesis, respectively. }
\end{figure}

Figure~\ref{fig:zh_inv_linv} (a) shows 95\% confidence level (CL)
limits on the production rate of an invisibly decaying Higgs boson
relative to the predicted SM Higgs production rate, defined as
$$\BR(\ho\to\X\X)
\frac{\sigma(\ee\to\Zo\ho)}
     {\sigma(\ee\to\Zo\HoSM)}
     =\BR(\ho\to\X\X)R_{\mathrm{\sigma}}$$
where $\sigma(\ee\to\Zo\ho)$ and $\sigma(\ee\to\Zo\HoSM)$ are the
production cross-sections of the invisibly decaying Higgs boson and
the SM Higgs boson, respectively, and $\BR(\ho\to\X\X)$ is the
branching ratio for the Higgs boson decay into a pair of invisible
particles.
                                   
The observed and expected ratios shown in the figure are obtained from
the results of this search, combined with LEP I data~\cite{LEP1-inv},
and with results from the \rmee\ and \mm\ channels of the decay-mode
independent searches~\cite{DECind}. For LEP I results, the recoil mass
information is used as a discriminating variable, incorporated using a
Gaussian mass resolution function; for the channels from the
decay-mode independent search, the distribution of the squared recoil
mass is used as a discriminant.

The full line in Figure~\ref{fig:zh_inv_linv} (a) represents the
observed upper limit at 95\% CL on the relative production rate as a function
of the Higgs boson mass.  A Higgs boson which couples to the
\Zo\ boson with SM strength and which decays exclusively into
invisible final states is excluded up to a mass of \limit~\gev\ at
95\% CL assuming $\BR(\ho\to\X\X)=100\%$, while a limit of
\limie\ \gev\ is expected. The
compatibility of the data with the expected background is quantified using
the confidence ({\it p}-value) for background-only hypothesis,
$1-\mathrm{CL_{b}}$ (see Ref.~\cite{Higgs-searches}) which is plotted
in Figure~\ref{fig:zh_inv_linv} (b).
\subsection{Limits on the production of nearly invisibly 
  decaying Higgs bosons\label{sec:inv_near}}
\begin{figure}[tbp]
  \centerline{
    \includegraphics[width=0.52\linewidth,clip]{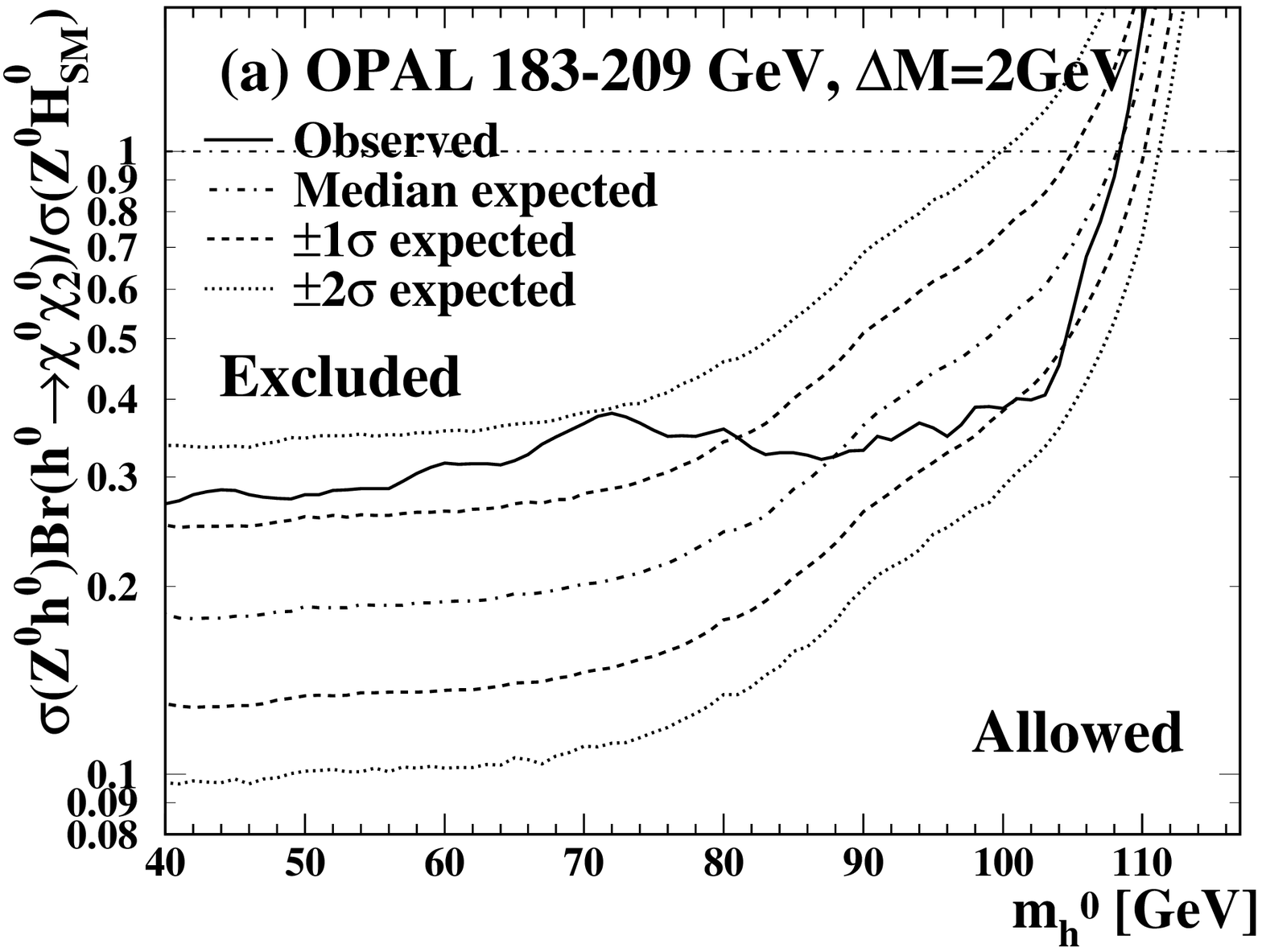}
    \includegraphics[width=0.52\linewidth,clip]{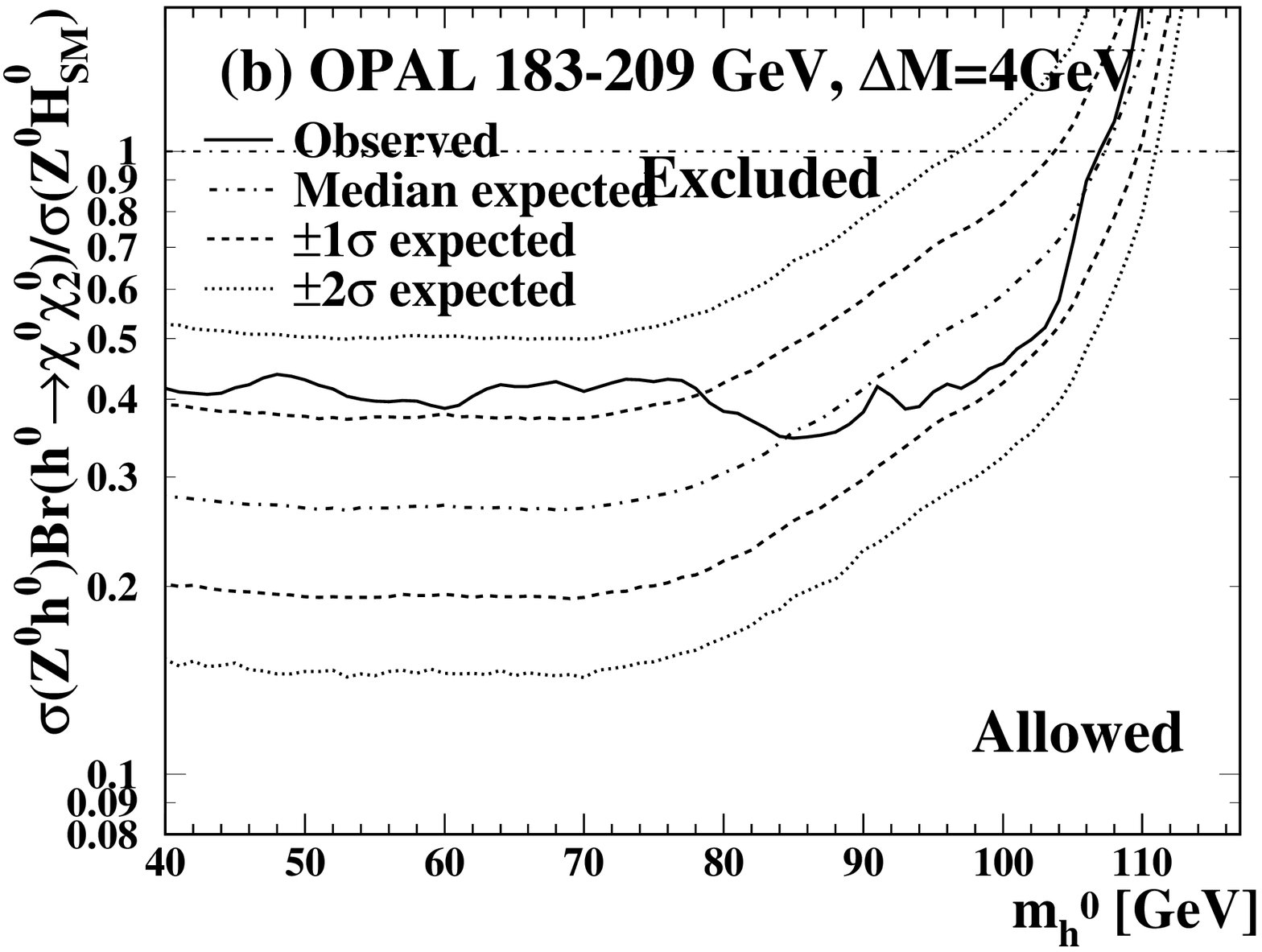}
  } \centerline{
    \includegraphics[width=0.52\linewidth,clip]{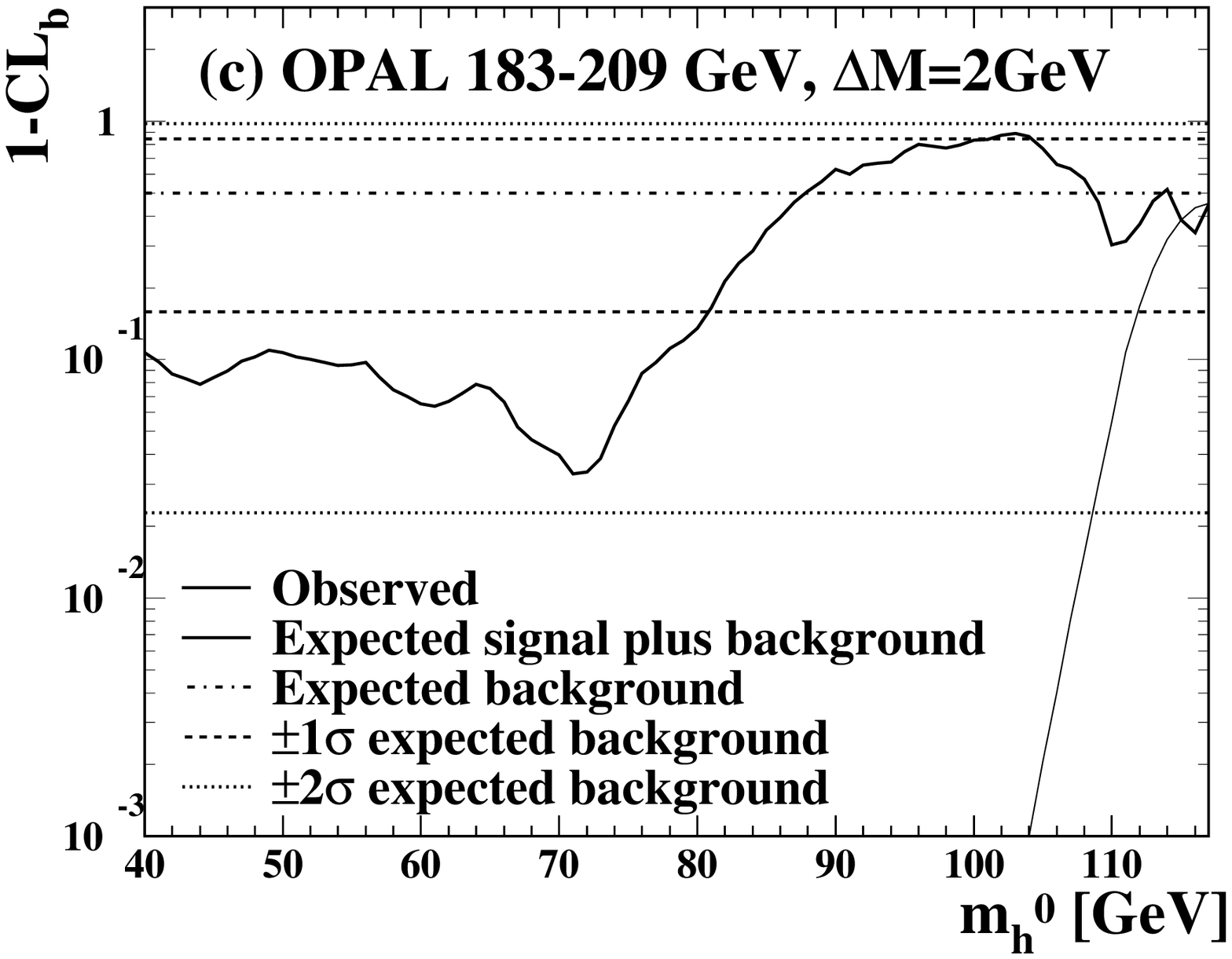}
    \includegraphics[width=0.52\linewidth,clip]{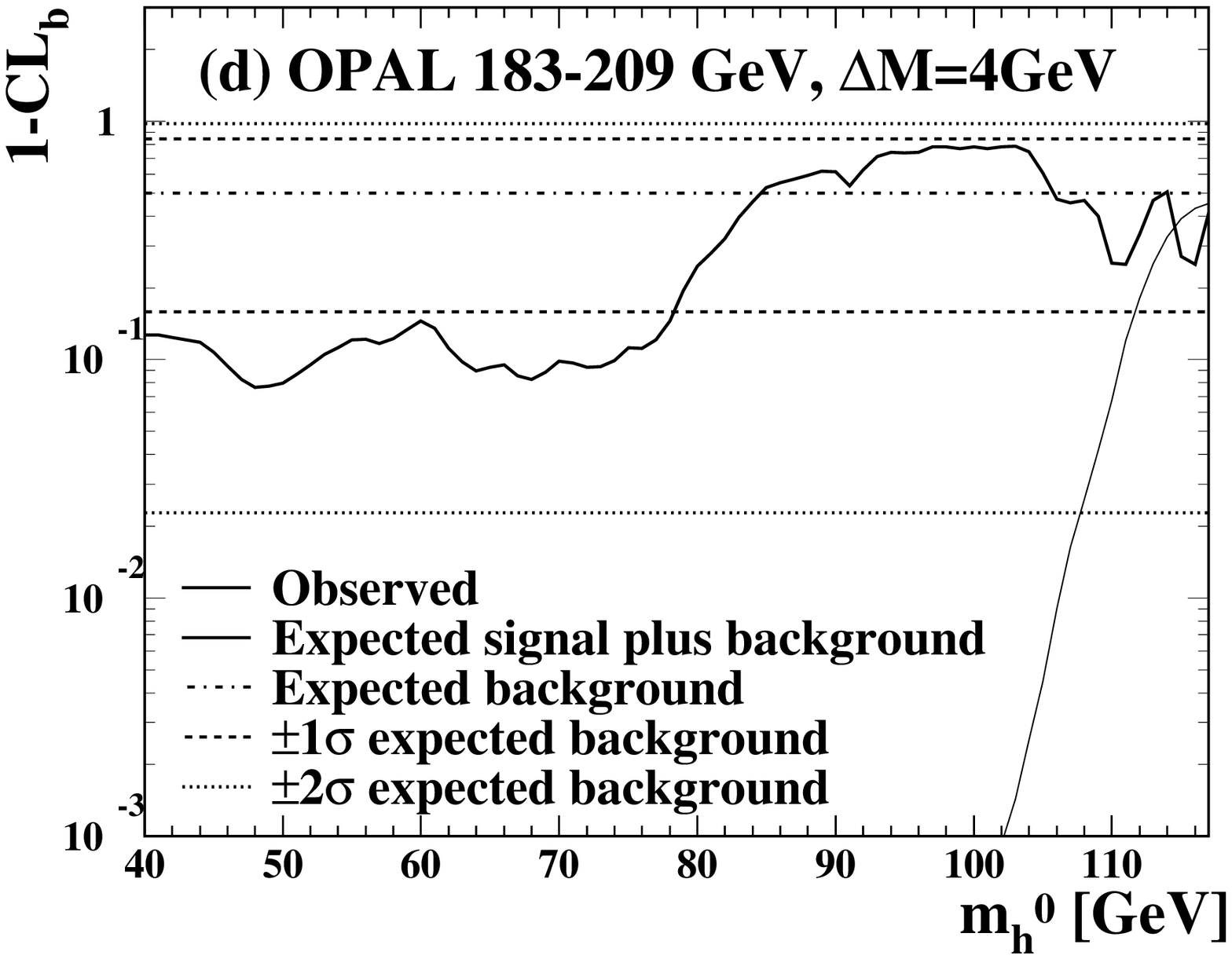}
  }
  \caption[]{\label{fig:zh_inv_lnear}\sl Limits on the relative
    production rate for $\rmee\to\Zo\ho\to\Zo\X\XD$ (nearly
    invisible decay) at the 95\% CL, normalised to the SM production rate
    for $\rmee\to\Zo\Ho$, (a) for $\dm=2$ \gev\ and (b) for $\dm=4$ \gev,
    assuming $\BR(\ho\to\X\XD)=100\%$ as a function of \mh. Figure~(c) and
    (d) show the \clb\ for $\dm=2$ and $4$~\gev, respectively.  }
\end{figure}

The results obtained from two nearly invisible decay modes
($\XD\to\X\Zs$ and $\XD\to\X\gamma$) in this analysis are combined at
each \dm, where the lower of the two efficiencies is taken as the
combined efficiency.  The limit calculation uses only results from
this analysis.  In Figure~\ref{fig:zh_inv_lnear} (a) and (b), limits
on the production rate for a nearly invisibly decaying Higgs boson
with $\dm=$ 2 and 4 \gev\ are shown for the data taken between
183~\gev\ and 209~\gev. The production rate of the nearly invisibly
decaying Higgs boson relative to the SM Higgs production rate is
defined as
$$\BR(\ho\to\X\XD)R_{\mathrm{\sigma}}$$ where $\BR(\ho\to\X\XD)$ is
the branching ratio for the decay into nearly invisible particles.
The dependence of $1-\mathrm{CL_{b}}$ on the Higgs mass for nearly
invisibly decaying Higgs bosons is shown in
Figure~\ref{fig:zh_inv_lnear} (c) and (d).  A Higgs boson coupling to
the \Zo\ boson with SM strength and decaying into the nearly invisible
final states is excluded up to a mass of \limna\ and \limnb~\gev\ at
95\% CL for $\dm=2$ and 4~\gev, respectively, assuming
$\BR(\ho\to\X\XD)=100\%$. The corresponding expected limits are
\limnae\ and \limnbe~\gev.
\section{Conclusion\label{sec:inv_conc}}
A search for invisibly decaying Higgs bosons has been performed using
the data collected by the OPAL experiment at centre-of-mass energies
between 183 and 209~\gev, corresponding to an integrated luminosity of
\lumtot~\pb. The search has not shown any excess over the expected
background from SM processes. Limits on the production of invisibly
decaying Higgs bosons were calculated combining the results with those
from LEP I and those from \rmee\ and \mm\ channels of a decay-mode
independent search at LEP II. Invisibly decaying Higgs bosons with
masses below \limit~\gev\ are excluded at $95\%$ CL if they are
produced with SM cross-sections, assuming $\BR(\ho\to\X\X)=100\%$. The
search criteria were also applied to a search for nearly invisibly
decaying Higgs bosons. Limits of \limna\ and \limnb~\gev\ were
obtained for $\dm=\mXD-\mX=2$ and 4~\gev, respectively.
\section*{Acknowledgements}
We particularly wish to thank the SL Division for the efficient operation
of the LEP accelerator at all energies
 and for their close cooperation with
our experimental group.  In addition to the support staff at our own
institutions we are pleased to acknowledge the  \\
Department of Energy, USA, \\
National Science Foundation, USA, \\
Particle Physics and Astronomy Research Council, UK, \\
Natural Sciences and Engineering Research Council, Canada, \\
Israel Science Foundation, administered by the Israel
Academy of Science and Humanities, \\
Benoziyo Center for High Energy Physics,\\
Japanese Ministry of Education, Culture, Sports, Science and
Technology (MEXT) and a grant under the MEXT International
Science Research Program,\\
Japanese Society for the Promotion of Science (JSPS),\\
German Israeli Bi-national Science Foundation (GIF), \\
Bundesministerium f\"ur Bildung und Forschung, Germany, \\
National Research Council of Canada, \\
Hungarian Foundation for Scientific Research, OTKA T-038240, 
and T-042864,\\
The NWO/NATO Fund for Scientific Research, the Netherlands.\\

\newpage

\end{document}